\documentclass[conference,10pt]{IEEEtran}
\usepackage{epsfig,rotating,setspace,latexsym,amsmath,epsf,amssymb,amsfonts,bm,theorem,subfigure,epstopdf}
\usepackage{cite,authblk}
\usepackage{bbm}
\usepackage{color}
\usepackage{mathtools}
\usepackage{algorithm}
\usepackage[noend]{algpseudocode}

\algrenewcommand\algorithmicforall{\textbf{foreach}}
\algrenewcommand\algorithmicindent{.8em}

\DeclarePairedDelimiter{\ceil}{\lceil}{\rceil}
\newtheorem{theorem}{Theorem}
\newtheorem{lemma}{Lemma}
\newtheorem{corollary}{Corollary}
\newtheorem{definition}{Definition}
\newenvironment{Proof}[1]{\medskip\par\noindent{\bf Proof:\,}\,#1}{{\mbox{\,$\blacksquare$}\par}}
\IEEEoverridecommandlockouts
\allowdisplaybreaks

\newcommand{\figref}[1]{\figurename~\ref{#1}}

\title{Group Testing with Non-identical \\ Infection Probabilities}
	
\author{Mustafa Doger \qquad Sennur Ulukus\\
\normalsize Department of Electrical and Computer Engineering\\
\normalsize University of Maryland, College Park, MD 20742\\
\normalsize  \emph{doger@umd.edu} \qquad \emph{ulukus@umd.edu}}

\begin{document}

\maketitle

\begin{abstract}
We consider a zero-error probabilistic group testing problem where individuals are defective independently but not with identical probabilities. We propose a greedy set formation method to build sets of individuals to be tested together. We develop an adaptive group testing algorithm that uses the proposed set formation method recursively. We prove  novel  upper  bounds on the number of tests for the proposed algorithm. Via numerical results, we show that our algorithm outperforms the state of the art, and performs close to the entropy lower bound.
\end{abstract}

\section{Introduction}
Introduced by Dorfman \cite{Dorfman1943} in 1943, group testing aims to find defective items in a population $\mathcal{P}$ using a minimum number of tests. The main idea is to mix samples from several individuals and test them all at once. A negative result implies that all items in the pool are defectless with only one test spent, whereas a positive result requires further tests for classification. Group testing is known to perform better than individual testing in sparse regimes, i.e., when the number of defectives is small compared to the size of the population.

Group testing algorithms are divided into two classes: Non-adaptive testing refers to algorithms where the tests to be performed are designed before testing starts, hence they can be run in parallel. Adaptive testing, on the other hand, allows the test designer to modify the test pools based on previous results \cite{gt-survey}. Group testing algorithms have been studied under various models, such as, combinatorial model, in which $k$ out of $n$ individuals are defective, where $k$ is a fixed deterministic number; and the i.i.d.~prior model in which each item is independently defective with identical probability $p$. The combinatorial model is studied in \cite{hwang-book} and a well-known optimal algorithm is given in \cite{hwang-gbsa}. The i.i.d.~prior model is studied in \cite{iid-model}. A survey of main group testing problems and well-known results can be found in \cite{gt-survey}.

Although the i.i.d.~prior case has been studied thoroughly in the literature, in many cases, it becomes a naive approach since infections are not identically distributed among different groups with different characteristics, such as, age, gender, family history, profession. Hence, we consider the independent and non-identically distributed (i.n-i.d.) model defined in \cite{jaggi-indep}, where the infection status of each individual is independent from other individuals, but not necessarily identically distributed. Our goal is to classify the individuals in this population according to their infection status using a minimum number of tests, and with zero error. 

\subsection{Related Works}
Our work is closely related to the adaptive group testing problem studied in \cite{jaggi-indep}, and then in \cite{bristol-indep}, for i.n-i.d.~probabilistic model. Our paper is also related to the adaptive group testing problem studied in \cite{hwang-gbsa} for the combinatorial model. In this paper, we build our ideas on the methods proposed by \cite{jaggi-indep} and \cite{hwang-gbsa}, and improve on the expected number of tests as well as on the upper bound given in \cite{jaggi-indep}. 

Reference \cite{hwang-gbsa} proposes an optimal adaptive group testing scheme for the combinatorial model. The algorithm modifies the traditional binary splitting algorithm (BSA) \cite{sobel-bsa} by first finding an optimal size to perform an initial test, and then recursively halves the population into two parts until a defective is found. The same procedure is then applied to the remaining population that has not been classified yet.

The method proposed in \cite{jaggi-indep} starts out by partitioning the population into disjoint sets using the probability of containing no defective elements. Each set size is determined by the maximum entropy (ME) principle, i.e., the probability of containing no defective elements is close to $1/2$. For positive test results, similar to binary splitting, sets are divided into two parts using either a Shannon-Fano-Huffman (SFH) or again an ME approach at each level to find all defectives in all sets.

The method proposed in \cite{bristol-indep} builds on the ideas proposed in \cite{jaggi-indep}. The algorithm partitions the population into disjoint bins containing items with similar probabilities of being defective. Then, items in bins are grouped together in sets such that the sum of their probabilities is close to $1/2$. For each set within each bin, an SFH tree is used to find the leftmost defective on the tree and all inconclusive right nodes of a set are grouped together and further tested to find all other defectives, one at a time, recursively until all items are classified.

\subsection{Our Contributions}
We improve the performance of the algorithm proposed in \cite{jaggi-indep} by proposing a new scheme for testing with a recursive set formation structure. We give novel upper bounds for the new algorithm. In our experiments, we show the performance of our algorithm together with the novel upper bounds, and also the performances of the algorithms in \cite{jaggi-indep} and \cite{bristol-indep}. Our algorithm starts out by grouping items that have the highest probability of being defective to form sets, and stops when the probability of containing no defective elements reaches $1/2$. We use SFH or ME trees to find a defective within each set. We then apply the same procedure for the remaining population that is not classified yet until all defectives are found.

Next, we list the differences of our algorithm from \cite{jaggi-indep, bristol-indep}; we elaborate on these differences in detail and give reasons why these differences result in an improved algorithm in Section~\ref{ssec:sf/huffman algo}. Specifically, we extend the set formation stage of \cite{jaggi-indep} by proposing an explicit, simple yet effective method, and prove that this method results in sets with mean less than $1$, which allows the application of information-theoretic results on SFH trees to these sets. While \cite{jaggi-indep} divides any set that is contaminated into two parts and tests them both, we never test the right child of any set/subset, which results in gains.

The method in \cite{bristol-indep} uses sums of probabilities to create a set (which should be about $1/2$), whereas we use products of probabilities. By drawing an analogy to data compression, ours performs better as explained in Section~\ref{ssec:sf/huffman algo}. The method in \cite{bristol-indep} uses bin structures that may potentially result in suboptimal groups depending on the probability distribution. Moreover, \cite{bristol-indep} combines inconclusive right nodes of a tree and tests them together, which is not optimal since these nodes do not form a group with an optimal size. We tackle this problem by putting inconclusive right nodes back into the population as in \cite{hwang-gbsa}. As a byproduct, our sets are not necessarily disjoint.

\section{System Model}
We consider a population, $\mathcal{P}=\{X_{1}, X_{2}, \ldots, X_{|\mathcal{P}|}\}$, where $X_{i}\in\{0,1\}$ represents the infection status of the $i$th individual and $\mathbb{P}(X_{i}=1)=p_{i}$. We denote the infection status vector as $\bm{X}=(X_{1}, X_{2}, \ldots, X_{|\mathcal{P}|})$ and the corresponding probability vector as $\bm{p}=(p_{1}, p_{2}, \ldots, p_{|\mathcal{P}|})$. Our recovery vector $\bm{Y}=(Y_{1}, Y_{2}, \ldots, Y_{|\mathcal{P}|})$ is necessarily equal to $\bm{X}$ as we require zero error recovery. The expected number of defective items in $\mathcal{P}$ is $\mu=\sum_{i=1}^{|\mathcal{P}|} p_{i}$. Without loss of generality, we assume $p_{i} < \frac{1}{2}$, as group testing does not offer any advantages otherwise.

We define a set of individuals to be tested as $\mathcal{S}$. We call a set contaminated if it contains at least one defective. For a set of individuals $\mathcal{S}$, $\mu_{\mathcal{S}}$ denotes the expected number of defectives in $\mathcal{S}$, i.e., $\mu_{\mathcal{S}}=\sum_{j \in \mathcal{S}} p_{j}$, and $P_{\mathcal{S}}$ denotes the probability that $\mathcal{S}$ is not contaminated, i.e., $P_{\mathcal{S}}=\prod_{j \in \mathcal{S}} (1-p_{j})$. As our adaptive algorithm gradually classifies items of $\mathcal{P}$, we denote the universal set of all yet-to-be-classified items as $\mathcal{P}^{*}$. If $\mathcal{A}$ is a subset of $\mathcal{S}$, then we denote the probability that $\mathcal{A}$ is contaminated given $\mathcal{S}$ is contaminated as $P_{\mathcal{A},\mathcal{S}}$, i.e.,
\begin{align}
    P_{\mathcal{A},\mathcal{S}}=\frac{1-\prod_{i\in \mathcal{A}}(1-p_{i})}{1-\prod_{i\in \mathcal{S}}(1-p_{i})}.
    \label{me-partition-eqn}
\end{align}

Obtaining a set with desired properties is one of our main goals in order to achieve well-performing algorithms. We define the following desired property of a set.

\begin{definition}
A set of items $\mathcal{S}$ is said to be saturated if 
\begin{align}
    P_{\mathcal{S}}=\prod_{i\in \mathcal{S}} (1-p_{i}) \leq \frac{1}{2}.
    \label{saturated eqn}
\end{align}
\end{definition}

Apart from $P_{\mathcal{S}} \leq \frac{1}{2}$, we also want $\mu_{\mathcal{S}} \leq 1$ to make sure that the set is not oversaturated.

\section{Algorithms and Analysis}
\subsection{SFH Coding Based Refined Laminar Algorithm}
\label{ssec:sf/huffman algo}
Our main contribution is providing an improved version of the adaptive laminar algorithm proposed in \cite{jaggi-indep}, and tightening the upper bound for the expected number of tests. We summarize our version of adaptive laminar algorithm as follows:
\begin{enumerate}
  \item \label{first item} We start by creating a saturated set of items $\mathcal{S}$ from $\mathcal{P}^{*}$ using \textit{Max-to-Min Greedy Saturation} described in Section~\ref{sssec:saturating a set}. We note that $\mathcal{P}^{*}=\mathcal{P}$ at the beginning.
  \item \label{second item} For this created set $\mathcal{S}$, we first perform a first stage/root test, i.e., we test all items in $\mathcal{S}$ together:
  \begin{enumerate}
      \item If $\mathcal{S}$ is not contaminated, the test result will be negative and we classify all items in $\mathcal{S}$ as defectless, so we return to step \ref{first item} to create a new set.
      \item If $\mathcal{S}$ is contaminated, we construct a SFH tree for the contaminated set $\mathcal{S}$. Then we set $\mathcal{C} := \mathcal{S}$ and continue with step \ref{third item}.
  \end{enumerate}
  \item \label{third item} If $|\mathcal{C}|=1$, we are done, i.e., we classify the single item in $\mathcal{C}$ as defective and return to step \ref{first item} to create a new set. If $|\mathcal{C}| > 1$, calling the left child of $\mathcal{C}$ as $\mathcal{L}$ and the right child as $\mathcal{R}$, we first test node $\mathcal{L}$:
    \begin{enumerate}
        \item If $\mathcal{L}$ is not contaminated, test result will be negative and we classify all items in $\mathcal{L}$ as defectless. Here, we already know $\mathcal{R}$ is contaminated since $\mathcal{L} \cup \mathcal{R} = \mathcal{C}$. Thus, we set $\mathcal{C} := \mathcal{R}$ and continue with step \ref{third item}.
        \item If $\mathcal{L}$ is contaminated, test result will be positive. In this case, we first put the items in $\mathcal{R}$ back into $\mathcal{P^{*}}$. Then we set $\mathcal{C} := \mathcal{L}$ and continue with step \ref{third item}.
  \end{enumerate}
\end{enumerate}

In short, we create a set $\mathcal{S}$ using \textit{Max-to-Min Greedy Saturation} and if $\mathcal{S}$ is not contaminated we spent only one test for the classification of the entire set. If $\mathcal{S}$ is contaminated, we find the leftmost defective item in SFH tree by repeatedly testing the left child of the leftmost contaminated nodes. Once we are done with a set, we create a new set until the whole population is classified. We note that the new set and the previous set might not be disjoint, since we might take some items from $\mathcal{P}^{*}$ that we put back to $\mathcal{P}^{*}$ while finding previous defective item. The improvement of our algorithm comes from the fact that we never test the right child of any node.

Although both \cite{jaggi-indep} and \cite{bristol-indep} propose similar algorithms, our version performs significantly better, and offers zero error testing. In comparison to \cite{jaggi-indep}, we form our sets using a specific method, \textit{Max-to-Min Greedy Saturation} which makes sure both $P_{\mathcal{S}} \leq \frac{1}{2}$ and $\mu_{\mathcal{S}} \leq 1$, while their rigorously proven algorithm requires $\frac{1}{2} \leq P_{\mathcal{S}} \leq \frac{3}{4}$, which is not as tight as desired. From an algorithmic perspective, we never waste tests on the right child of any node and our sets are not necessarily disjoint, which explains our considerable gains in performance.

Reference \cite{bristol-indep} suggests to build sets with mean, $\mu_{\mathcal{S}}$, close to $1/2$ whereas our suggestion is to build sets where probability of containing no defective, $P_{\mathcal{S}}$, is close to $1/2$. Intuitively, if we consider each test result as a way of compressing data into $1$s and $0$s, computing the mean will add a weight for each defective item in a population, however, a positive test result cannot differentiate between one or many defectives. Therefore, using $P_{\mathcal{S}}\approx \frac{1}{2}$ is a better approach than $\mu_{\mathcal{S}}\approx \frac{1}{2}$ for the purpose of entropy maximization.

Moreover, in \cite{bristol-indep}, each created set is disjoint and all elements of inconclusive right nodes of a set are combined and tested at once. Although this is a better approach than testing both children as in \cite{jaggi-indep}, we note that half of the time, the size of the combined inconclusive nodes is less than half the size of an optimum group. Building contiguous groups --which are not disjoint-- with a recursive algorithm, where inconclusive right nodes are put back into the population, tackles this problem, which is reminiscent of Hwang's generalized BSA (HGBSA) approach given in \cite{hwang-gbsa}. As an analogy, just like Hwang's $(n,m)$ problem is reduced to $(n',m')$ problem at each step, our method reduces $(\mathcal{P},\bm{p})$ problem to $(\mathcal{P}^{*},\bm{p}^{*})$ problem.

We also note that we do not require any constraints on the distribution of $p_{i}$'s such as bin structures of \cite{bristol-indep} and only need a simple greedy algorithm to build a set that can potentially combine an item with any probability in the range $[0,\frac{1}{2}]$.

\subsection{ME Based Refined Laminar Algorithm}
In the ME based algorithm, instead of using an SFH tree to partition contaminated sets/subsets, we use the following approach: at step \ref{third item} of laminar algorithm described in Section~\ref{ssec:sf/huffman algo}, we minimize $\left|P_{\mathcal{L},\mathcal{C}}-\frac{1}{2}\right|$ to partition $\mathcal{C}$ into $\mathcal{L}$ and $\mathcal{R}$ as proposed in \cite{jaggi-indep}. Note that, although SFH and ME based algorithms partition nodes with different methods, they both avoid testing right children of the nodes and result in considerable gain in terms of number of tests used.

Here, an important remark is, neglecting quantization errors, this partition results in the conditional probability of the left child being contaminated is close to $\frac{1}{2}$, whereas the same probability for the right child is higher than $\frac{1}{2}$. Hence, we always create the set for the left child with entropy maximization, and test this near-optimal left child instead of the right child.

\subsection{Example Run}
\begin{figure}[t]
	\centerline{\includegraphics[width=0.65\columnwidth]{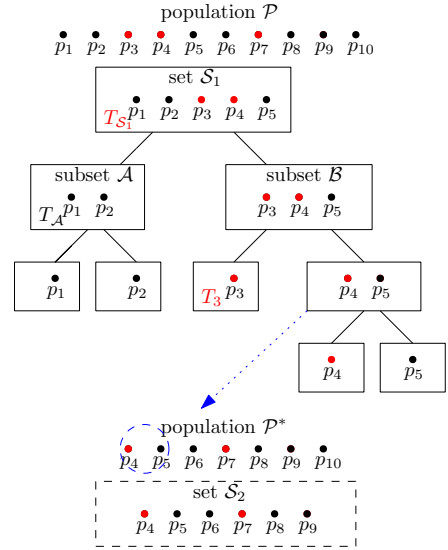}}
	\caption{Example step of a testing procedure for the SFH construction.}
	\label{huf-figure} 
\end{figure}

In \figref{huf-figure}, we show an example run for the SFH construction on a sample population $\mathcal{P}$, where red dots represent defective items and black dots represent defectless ones. A set $\mathcal{S}_{1}$ that satisfies the required properties is created and tested. Since $\mathcal{S}_{1}$ is contaminated, the test result, $T_{\mathcal{S}_{1}}$, will be positive. As a result, we build an SFH tree using given probabilities and test the left child of $\mathcal{S}_{1}$, i.e., $\mathcal{A}$. Since $T_{\mathcal{A}}$ will be negative, we classify all items in $\mathcal{A}$ as defectless. Here, note that although SFH tree determined further partitioning of $\mathcal{A}$ into $p_{1}$ and $p_{2}$, we do not need this partitioning as a result of $T_{\mathcal{A}}$. Also, from this result, we directly conclude that $\mathcal{B}$ is contaminated and we test the left child of $\mathcal{B}$. The result, $T_{3}$, will be positive, and we classify item $p_{3}$ as defective, and put the right child of $\mathcal{B}$ back into $\mathcal{P}$. From here on, we create a new set, $\mathcal{S}_{2}$, to find the next defective item in $\mathcal{P^{*}}$. As we emphasized earlier, in this scenario, $\mathcal{S}_{1} \cap \mathcal{S}_{2} \neq \emptyset$, i.e., they are not disjoint.

A similar procedure for this sample population would apply in case of the ME algorithm, however note that the partitioning resulting from ME principle does not necessarily have to have the same branching structure as the SFH case.

\subsection{Saturating a Set $\mathcal{S}$} \label{sssec:saturating a set}
Saturated sets are desired since saturation makes sure that sets have sufficiently many items so that group testing may perform better than individual testing. At the same time, a set with $\mu_{\mathcal{S}}\leq1$ is desired, because we do not want the length of the branch for $p_{i}$'s to be more than $\log_{2}\frac{1}{p_{i}} +1$.  Thus, one should avoid overfilling the set, which results in a waste of tests since the first stage test result will be positive with high probability for an overfilled set. Further, since our algorithm is recursive, we need a low-cost method. A simple greedy algorithm satisfying these properties is explained next.

\subsubsection{Max-to-Min Greedy Saturation} \label{sssec:sat}
The algorithm starts filling the set $\mathcal{S}$ by first adding the item with $\max_{i \in \mathcal{P^{*}}}  p_{i}$ from the unclassified population $\mathcal{P^{*}}$. In the later stages, the algorithm adds the next item with maximum $p_{i}$ from $\mathcal{P^{*}}$ and repeats this until the set $\mathcal{S}$ is \emph{saturated}, i.e., $P_{\mathcal{S}}=\prod_{i\in \mathcal{S}} (1-p_{i}) \leq \frac{1}{2}$. Once the set $\mathcal{S}$ is saturated the algorithm stops. If there are no unclassified items left in $\mathcal{P^{*}}$ to saturate $\mathcal{S}$, i.e., $\mathcal{P^{*}}=\emptyset$, the algorithm stops and returns the unsaturated set.

\begin{lemma}
Consider a saturated set $\mathcal{S}$ created by using Max-to-Min Greedy Saturation. Let $d_{m}$ denote the item with maximum $p_{i}$ in $\mathcal{S}$ and let $p_{m}$ denote its probability. Let $d_{l}$ denote the item with minimum $p_{i}$ in $\mathcal{S}$ and let $p_{l}$ denote its probability. Then, the saturated set $\mathcal{S}$ satisfies the following condition,
 \begin{align}
     \mu_{\mathcal{S}} < 1 -(p_{m}-p_{l}).
     \label{complete eqn}
 \end{align}
\end{lemma}

\begin{Proof}
Let us denote $\mathcal{S}\setminus\{d_{l}\}$ as $\mathcal{S^{*}}$. Since $\mathcal{S^{*}}$ is not saturated, $P_{\mathcal{S}^{*}}=\prod_{i\in \mathcal{S}^{*}} (1-p_{i}) > \frac{1}{2}$, which can be rewritten as,
\begin{align}
    2 &> \prod_{i\in \mathcal{S}^{*}}\frac{1}{1-p_{i}} \\
    &= \left( \prod_{i\in \mathcal{S}^{*}\setminus\{d_{m}\}} \frac{1}{1-p_{i}} \right) \frac{1}{1-p_{m}}.
\end{align}
By reordering terms, we get
\begin{align}
    2-2p_{m} &>  \prod_{i\in \mathcal{S}^{*}\setminus\{d_{m}\}} \frac{1}{1-p_{i}} \\
    &= \prod_{i\in \mathcal{S}^{*}\setminus\{d_{m}\}}  
    \left( \sum_{j=0}^{\infty} p_{i}^{j} \right) \label{geosum} \\ 
    &\geq 1+ \sum_{i\in \mathcal{S}^{*}\setminus\{d_{m}\}} p_{i}, \label{weierstrass}
\end{align}
where we use geometric sum expansion in \eqref{geosum}, and Weierstrass product inequality in \eqref{weierstrass}. By adding $p_{m}$ to both sides, and subtracting $1$ from both sides, we get
\begin{align}
     1-p_{m} >  \sum_{i\in \mathcal{S}\setminus\{d_{l}\}} p_{i}. \label{incompleteset}
\end{align}
After adding $d_{l}$, the set becomes saturated. Hence,  adding $p_{l}$ to both sides of \eqref{incompleteset}, we get \eqref{complete eqn}, concluding the proof.
\end{Proof}

We note that $\mu_{\mathcal{S}} \leq 1$ does not hold for every saturated set, however, our Max-to-Min Greedy Saturation guarantees \eqref{complete eqn}. Stopping right after the set is saturated not only guarantees \eqref{complete eqn}, but also ensures that we do not overfill the set $\mathcal{S}$. Hence, the entropy of the set is near maximum.

There are several important points to note about the SFH construction. Branch lengths in the SFH tree, i.e., $\ceil*{\log_{2}\frac{1}{p_{i}}}$, are based on the number of edges on a branch. When we count the number of tests, we need to add a $+1$ to the SFH branch lengths, since we are counting the number of nodes on a branch, which represent the number of tests. Another important fact is, we do not create SFH trees for uncontaminated sets, and for the contaminated sets, we only find the leftmost defective item on an SFH tree, and individuals on the right side, some of whom are potentially defective, will be put back into $\mathcal{P}^{*}$ during this process. As a result, the number of times we create an SFH tree is equal to the number of defectives in our population. The fact that $\mu$ is equal to the expected number of SFH construction is a good indicator of the  computational cost before testing starts. Finally, as probabilities are sorted in \textit{Max-to-Min Greedy Saturation} stage, using arithmetic codes, such as Shannon coding, can reduce the complexity.

\begin{corollary} \label{sf cor}
A defective item with probability $p_{i}$ in a set $\mathcal{S}$ with  $\mu_{\mathcal{S}} \leq 1$ can be found in less than $\log_{2} \frac{1}{p_{i}} + 2$ tests using the SFH coding based algorithm.
\end{corollary}

\begin{Proof}
We first normalize $p_{i}$ and denote new probabilities as $\bar{p}_{i}= \frac{p_{i}}{\mu_{\mathcal{S}}}$. Using the SFH construction, the length of the branch $l_{i}$ associated with $\bar p_{i}$ satisfies
\begin{align}
    l_{i} & \leq \ceil*{\log_{2}\frac{1}{\bar{p}_{i}}} \\
    &< \log_{2}\frac{1}{p_{i}} + \log_{2}{\mu_{\mathcal{S}}} +1 \\
    &< \log_{2}\frac{1}{p_{i}} +1, \label{sf-tree}
\end{align}
where \eqref{sf-tree} follows from $\mu_{\mathcal{S}} < 1$. Accounting for the first stage test to be performed on $\mathcal{S}$ at the root of the SFH tree, the total number of tests to find a defective item is $l_{i}+1$.
\end{Proof}

Although the bound in \eqref{sf-tree} is not necessarily valid for the ME based algorithm, as stated in \cite{jaggi-indep}, if we neglect quantization errors while finding the minimizing $\left|P_{\mathcal{C}^{lk},\mathcal{P}^{k}}-\frac{1}{2}\right|$, where $\mathcal{C}^{lk}$ represents the left child set at stage $k$ and $\mathcal{P}^{k}$ the parent set at stage $k$, i.e., assume $P_{\mathcal{C}^{lk},\mathcal{P}^{k}}=\frac{1}{2}$, for all $k$, the same argument can be made for branches of the ME based partitioning,
\begin{align}
    1-\frac{1}{2^{k-1}+1} &= P_{\mathcal{C}^{rk}} \leq (1-p_{i}), \label{me-k} 
\end{align}
where $\mathcal{C}^{rk}$ represents the rightmost child of the tree at stage $k$. By rearranging the terms and replacing $k$ with $l_{i}$, we obtain
\begin{align}
    l_{i} \leq \ceil*{\log_{2}\frac{1}{p_{i}}} +1.
\end{align}
Here we use the rightmost child of the tree which results in a larger bound for $l_{i}$ than the left children.

\section{Bounds}
\subsection{Groups with No Defective Items} \label{ss:ENG}
When we create saturated sets, with probability $P_{\mathcal{S}}$, a set will be uncontaminated and we will not need to create any tree or partition the set into smaller subsets. When we bound the expected number of tests, we have to account for the tests spent on these uncontaminated sets. A defectless item can get discarded while testing a contaminated set or within an uncontaminated set. In the former case, we do not need to double count this test since we will count it when counting the number of tests spent to find defectives. In the latter case, we need to count these tests, however, tracing their probability is a complicated task. Nevertheless, we can bound the maximum number of uncontaminated sets in a deterministic manner, by the number that would be obtained if every item in the population were defectless. In this case, the number of tests spent on the first stage for negative groups will be bounded by the maximum number of disjoint partitions of $\mathcal{P}$.

\begin{lemma} \label{lemma2}
The maximum number of disjoint partitions of a population $\mathcal{P}$, where partitions are created using Max-to-Min Greedy Saturation, is less than $2\mu+1$.
\end{lemma}

\begin{Proof}
For a saturated set $\mathcal{S}$, from Weierstrass product inequality, we have 
\begin{align}
    \frac{1}{2} &\geq \prod_{i\in \mathcal{S}} (1-p_{i})\\
    &\geq 1-\sum_{i\in \mathcal{S}} p_{i}.
\end{align}
As a result, $\mu_\mathcal{S} \geq \frac{1}{2}$ for saturated sets. Thus,
\begin{align}
    \mu &= \sum_{\mathcal{S}_{j} \in \mathcal{P}} \mu_{\mathcal{S}_{j}}\\
    &\geq \sum_{\mathcal{S}_{j} \in \mathcal{P}} \frac{1}{2}.
\end{align}
This implies that there will be at most $2\mu$ disjoint saturated  sets. Counting the last set, which might be unsaturated, explicitly as $+1$, we get $2\mu+1$ as an upper bound.
\end{Proof}

We can further tighten the bound in the i.i.d.~case as follows.

\begin{lemma} \label{lemma3}
If each item is defective i.i.d.~with probability $p$ in a countably infinite population $\mathcal{P}$, and if a saturated set $\mathcal{S}$ has size $n$ satisfying $P_{\mathcal{S}}=(1-p)^{n} \leq 1/2$, then the expected number of items that have to be classified to encounter the first uncontaminated set is given by
\begin{align}
    \mathbb{E}\left[N_{n}\right]=\frac{(1-p)^{-n}-1}{p}.
    \label{iid-en}
\end{align}
\end{lemma}

\begin{Proof}
We start by numbering the items in $\mathcal{P}$, $1$ through $\infty$. Assume that the first consecutive $n$ defectless items have numbers $m+1$ to $m+n$. We will now explain, without loss of generality, that we can classify items in the same order as their numbers with our algorithm, hence we only need to find $\mathbb{E}[m+n]$, which we will call $\mathbb{E}\left[N_{n}\right]$. Since all elements in $\mathcal{P}$ have the same probability, we choose items according to their assigned numbers in ascending order to create saturated sets. For contaminated sets, while designing the branches of the SFH tree, again, we can order them from left to right in ascending order on leaves of the tree. As a result of this, while finding defective items from $1$ to $m$, all defectless items with a number less than $m$ will be discarded before we encounter an uncontaminated set. Once we are done with items from $1$ to $m$, the next set will contain items $m+1$ through $m+n$, which will be uncontaminated. As an analogy, the question at hand is the same as the expected number of coin tosses needed to encounter $n$ consecutive heads where probability of heads is $1-p$. By conditioning on the first toss result, we have
\begin{align}
    \mathbb{E}\left[N_{n}\right] =& \mathbb{E}\left[N_{n}|C_{1}=T\right]\mathbb{P}(C_{1}=T) \nonumber \\ 
    &+ \mathbb{E}\left[N_{n}|C_{1}=H\right]\mathbb{P}(C_{1}=H) \\
    =& \left(1+\mathbb{E}\left[N_{n}\right]\right)p+\mathbb{E}\left[N_{n}|C_{1}=H\right](1-p).
\end{align}
By iterating the conditioning on further coin toss results,
\begin{align}
    \mathbb{E}\left[N_{n}\right] &= n(1-p)^{n} + \sum_{i=0}^{n-1} (1-p)^{i}p\left(i+1+\mathbb{E}\left[N_{n}\right]\right).
\end{align}
Solving the above equation for $\mathbb{E}\left[N_{n}\right]$, we obtain \eqref{iid-en}. For illustrative purposes, one can use a Markov chain with states $0$ through $n$ where the probability of moving one state right is $(1-p)$, and with probability $p$ each state moves back to $0$. From this Markov chain perspective, the question becomes: starting from state $0$, what is the expected number of steps to reach state $n$, the answer to which is \eqref{iid-en}.
\end{Proof}

\begin{corollary} \label{cor2}
In the i.i.d.~case, the expected number of groups with no defective items, denoted as $\mathbb{E}_{TN}$, is less that $\mu+1$.
\end{corollary}

\begin{Proof}
Denote each individual's probability of being defective as $p$ and the size of each saturated set as $n$. For each saturated set the following is true,
\begin{align}
    P_{\mathcal{S}}=(1-p)^{n} \leq 1/2. \label{iid half bound}
\end{align}
Note that the expected number of people needed to encounter one negative group was given in \eqref{iid-en}, which is a stopping time. Thus, using Wald's identity, we obtain a bound on the expected number of groups with no defective items as follows
\begin{align}
    \mathbb{E}_{TN} &< \frac{|\mathcal{P}|}{\mathbb{E}\left[N_{n}\right]}+1 \\
    &= \frac{\mu}{(1-p)^{-n}-1}+1 \\
    &\leq \mu+1, \label{iid-etn}
\end{align}
where we count the last test explicitly as the set might not be saturated, and \eqref{iid-etn} follows from \eqref{iid half bound}.
\end{Proof}

We can use this numbering idea for the i.n-i.d.~case as well if we use Shannon coding. However, in this case we cannot use a Markov chain with static probabilities, as discussed at the end of proof of Lemma~\ref{lemma3}, since each time we move back to state $0$, transition probabilities to other states will change. Thus, even though the upper bound we obtain for the i.i.d.~case is $\mu+1$, the upper bound we obtain for the i.n-i.d.~case is only $2\mu+1$. We finally note that, we experimentally observe that $\mu+1$ still approximately holds for the i.n-i.d.~case as well, even though the bound we can rigorously prove is $2\mu+1$. 

\subsection{Upper Bound on the Total Number of Tests}
First, we restate the upper bounds developed in \cite{jaggi-indep} and \cite{bristol-indep}. Then, we present the upper bound we develop for our algorithm in Theorem~\ref{our-bound}. The upper bound for the expected number of tests to recover all defective items for the laminar algorithm in \cite{jaggi-indep} is $\mathbb{E}[T]\leq 2H(\bm{X})+6\mu$, and the upper bound for the algorithm in \cite{bristol-indep} is $\mathbb{E}[T]\leq H(\bm{X})+4\mu + 2\sqrt{\mu(-\log_{2}(2\theta))}+1$. Note that $H(\bm{X})$ term of \cite{jaggi-indep} is multiplied by $2$, as they test each child of the contaminated sets/subsets. Also, although the $H(\bm{X})$ term of \cite{bristol-indep} has the same factor of $1$ as our result in Theorem~\ref{our-bound}, the bound in \cite{bristol-indep} is larger due to the additive term.

\begin{theorem}\label{our-bound}
Given a population $\mathcal{P}$ where item $i$ is defective with probability $p_{i}$, the expected number of tests to recover all defective items using our version of laminar algorithm based on an SFH construction can be bounded by,
\begin{align}
    \mathbb{E}[T] < H(\bm{X})+ 3\mu +1.
\end{align}
For the i.i.d.~case, this bound can be further tightened to,
\begin{align}
    \mathbb{E}[T] < H(\bm{X})+ 2\mu +1.
\end{align}
\end{theorem}

\begin{Proof}
Corollary~\ref{sf cor} shows each item with $p_{i}$ will be recovered in less than $\log_{2} \frac{1}{p_{i}}+2$ tests if they are defective. If we add the maximum number of disjoint partitions from Lemma~\ref{lemma2} to bound the tests spent on uncontaminated sets, we get
\begin{align}
    \mathbb{E}[T] < &\sum_{i \in \mathcal{P}} \left( p_{i} \log_{2}\frac{1}{ p_{i}} +2p_{i}\right) + 2\mu +1\\
    \leq &H(\bm{X})+ 3\mu +1, \label{p-log}
\end{align}
where \eqref{p-log} follows from the fact that, for $p_{i} \leq \frac{1}{2}$,  $p_{i}$ is less than $(1-p_{i})\log_{2}\frac{1}{(1-p_i)}$. For the i.i.d.~case, we can use Corollary~\ref{cor2} instead of Lemma~\ref{lemma2} to obtain the desired bound.
\end{Proof}

Note that our discussion in Section~\ref{sssec:saturating a set} gives an insight that Theorem~\ref{our-bound} would hold for ME based algorithm as well if quantization errors are neglected.

\section{Numerical Results}
\begin{figure}[t]
   \centerline{\includegraphics[width=0.97\columnwidth]{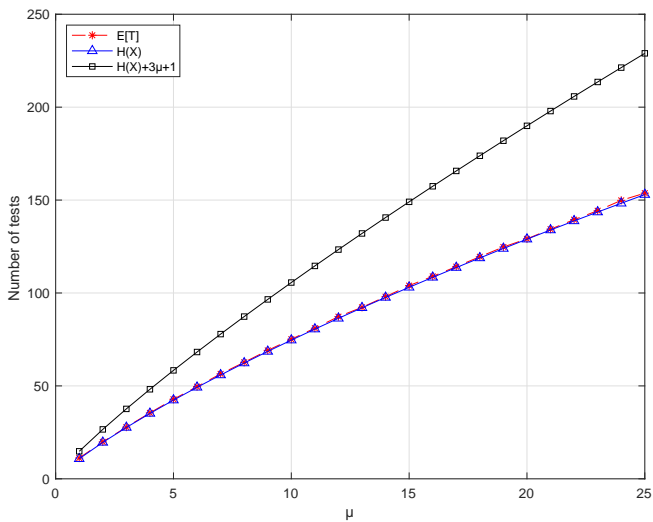}}
   \caption{Expected number of tests and theoretical bounds versus $\mu$.}
   \label{bounds1}
\end{figure}

In this section, we provide two experimental results. In the first experiment, we only consider the rigorously proven SFH coding based algorithm and its bounds. The population size is $|\mathcal{P}|=1000$. The infection probabilities, $\bm{p}$, follow a Dirichlet distribution. The parameter of Dirichlet distribution, $\alpha$, is chosen as $1$. There are in total $25$ different entropy points and the number of trials is $10000$ for each point. As seen from the curves in \figref{bounds1}, our algorithm  runs tightly on the entropy lower bound and on average there is less than $1$ test difference with the lower bound for the given region. Another important remark is, there is  room for improvement on the upper bound as we bounded the number of negative first stage groups in a deterministic manner with $2\mu+1$. We believe that this number should be around $\mu+1$ as we proved for the i.i.d.~case.

We run the second experiment with different conditions. Our aim is to compare the five different methods: ME based algorithm and SFH based algorithm proposed in this paper, laminar algorithm proposed in \cite{bristol-indep} (Kealy et. al.), ME based laminar algorithm proposed in \cite{jaggi-indep} (Li et. al.), and a very simple improved version of the same algorithm in \cite{jaggi-indep} (Li et. al. improved). This simple improvement is as follows: In \cite{jaggi-indep}, if a set (or subset) is contaminated, no matter what the result of the test for the left child is, the algorithm tests both the left and right children which is not necessary if the left child is not contaminated. Essentially, this can be seen as a correction on decoding rather than an improvement on the algorithm. 

In this second experiment, the population size is $|\mathcal{P}|=500$. The infection probabilities, $\bm{p}$, follow an exponential distribution truncated to $[0, 0.5]$. We choose $\frac{1}{\lambda}$ as $25$ linearly spaced values between $[0.0025, 0.025]$. The number of trials for each $\frac{1}{\lambda}$ is again $10000$. For the method in \cite{bristol-indep}, we choose $\theta=10^{-5}$, which has very little impact in any case, since as mentioned in \cite{bristol-indep}, the performance is insensitive to $\theta$. For the sake of neatness, we do not show any lower or upper bounds, but we note that our method is again within $1$ test of the entropy lower bound on average. 

There are a few things to note here. First, our SFH based method is slightly better than our ME based method. Second, although the method in \cite{jaggi-indep} has the worst performance among proposed methods, with a simple correction, it catches \cite{bristol-indep} immediately, even though \cite{bristol-indep} never tests the right children of a node, which should have given a good advantage to it. We believe that there are two reasons for this: First, as mentioned earlier, \cite{bristol-indep} uses $\mu_{\mathcal{S}} \approx 1/2$ which is not optimal, whereas \cite{jaggi-indep} uses $P_{\mathcal{S}} \approx 1/2$ similar to our method. Second, inconclusive right nodes in \cite{bristol-indep} have suboptimal sizes.

\begin{figure}[t]
   \centerline{\includegraphics[width=0.97\columnwidth]{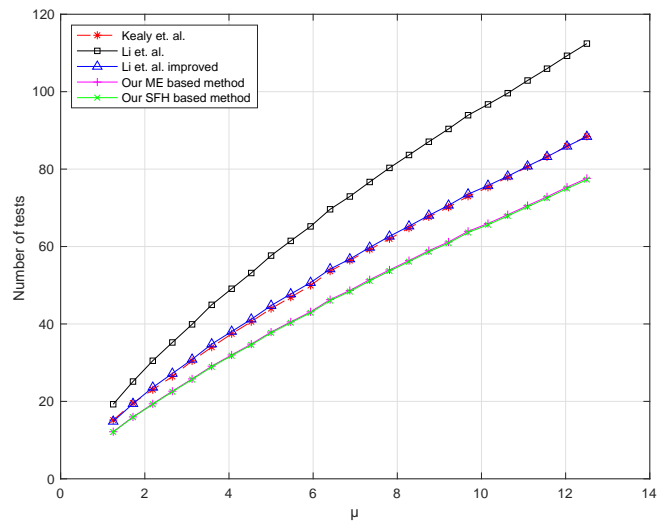}}
   \caption{Expected number of tests comparison between proposed methods.}
   \label{bounds2}
\end{figure}

\bibliographystyle{unsrt}
\bibliography{lib}

\begin{thebibliography}{1}

\bibitem{Dorfman1943}
R.~Dorfman.
\newblock The detection of defective members of large populations.
\newblock {\em Annals of Mathematical Statistics}, 14(4):436--440, December
  1943.

\bibitem{gt-survey}
M.~Aldridge, O.~Johnson, and J.~Scarlett.
\newblock Group testing: An information theory perspective.
\newblock {\em Foundations and Trends in Communications and Information
  Theory}, 15(3-4):196--392, December 2019.

\bibitem{hwang-book}
D.-Z. Du and F.~K. Hwang.
\newblock {\em Combinatorial Group Testing And Its Applications}.
\newblock World Scientific, 2 edition, December 1999.

\bibitem{hwang-gbsa}
F.~K. Hwang.
\newblock A method for detecting all defective members in a population by group
  testing.
\newblock {\em Journal of the American Statistical Association},
  67(339):605--608, September 1972.

\bibitem{iid-model}
J.~Wolf.
\newblock Born again group testing: Multiaccess communications.
\newblock {\em IEEE Transactions on Information Theory}, 31(2):185--191, March
  1985.

\bibitem{jaggi-indep}
T.~Li, C.~L. Chan, W.~Huang, T.~Kaced, and S.~Jaggi.
\newblock Group testing with prior statistics.
\newblock In {\em IEEE International Symposium on Information Theory}, pages
  2346--2350, June 2014.

\bibitem{bristol-indep}
T.~Kealy, O.~Johnson, and R.~Piechocki.
\newblock The capacity of non-identical adaptive group testing.
\newblock In {\em Allerton Conference on Communication, Control, and
  Computing}, pages 101--108, October 2014.

\bibitem{sobel-bsa}
M.~Sobel and P.~A. Groll.
\newblock Group testing to eliminate efficiently all defectives in a binomial
  sample.
\newblock {\em The Bell System Technical Journal}, 38(5):1179--1252, September
  1959.

\end{thebibliography}

\end{document}